\documentstyle[twocolumn,aps,prl,epsf,floats]{revtex}
\input{epsf}
\begin{document}
%\bibliographysty{prsty}
%\tighten
\title{A New Quantum Phase Transition in the Coupled Quantum Dots System }
\author{Natan Andrei$^1$, Gergely T. Zim\'anyi$^2$, and Gerd Sch\"on$^3$}
\address{$^{1)}$ Physics Department, Rutgers University, Piscataway, NJ 08855}
\address{$^{2)}$ Physics Department, University of California, Davis, CA
95616}
\address{$^{3)}$ Institut f\"ur Theoretische Festk\"orperphysik, 
Universit\"at  Karlsruhe, D-76128 Karlsruhe, Germany}
%\date{\today}
\address{\mbox{ }}
%\begin{abstract}
\address{\parbox{14cm}{\rm \mbox{ }\mbox{ }
We study two quantum dots in the limit of strong dot-lead coupling
and weak dot-dot tunneling. The model maps on Ising-coupled Kondo impurities.
We argue that a new quantum critical fixed point exists at
an intermediate value of the mutual capacitance, supporting non-Fermi liquid
behaviour. We construct the total conductance across the double dot
structure. It exhibits a strongly peaked behaviour as a function of the 
mutual capacitance, gate voltage, and temperature. 
}}
\address{\mbox{ }}
\address{\parbox{14cm}{\rm \mbox{ }\mbox{ }
}}
\address{\mbox{ }}
\maketitle
 
%\end{abstract}
 
\narrowtext
Electron tunneling through quantum dots is fundamentally affected by intriguing
many body effects. The Coulomb interaction imposes a prohibitive energy cost 
$E_{C}$ on the transfer of electrons, known as Coulomb 
blockade\cite{likharev,grabert1}. Fine tuning of the gate voltage $V_{\rm G}$
is required to reinstate charge flow, manifesting itself in sharp
conductance peaks as a function of $V_{\rm G}$.
%Suitable tuning of the gate voltage $V_{\rm G}$ can make
%the $N$ and $N+1$ electron states of the dot degenerate, 
%allowing the uninhibited flow of charges.
%This manifests itself in sharp conductance peaks as a function of $V_{\rm G}$
%, which have been confirmed in experiments\cite{grabert}.

Remarkably, the charge transfer is also accompanied by an orthogonality 
catastrophy. The analogy to the Kondo problem was recognized early
\cite{guinea}, with an exact formulation due to Matveev\cite{matveev}. For a 
single dot the Kondo type slow rearrangement of the electron states leads to 
a substantial downward renormalization of $E_{C}$\cite{bulgadaev}, as well as 
a smoothing of the conductance peaks\cite{schon,joyez}.
%The quantitative agreement between recent calculations\cite{schon} 
%and experiments\cite{joyez} lends further support to this picture.
Additional processes, such as the effect of higher order terms\cite{grabert2},
inelastic cotunneling\cite{nazarov}, and mapping to the 
out of equilibrium Anderson model\cite{wingreen} were also analyzed.

New effects arise when two such systems are allowed to interact.
We argue that a genuinely new and robust quantum phase transition takes 
place in the coupled dot system when their mutual capacitance is varied. 
It is driven by a change of the degeneracy of the ground 
state. Interdot tunneling will be included perturbatively.
It turns out to be a relevant operator, manifesting itself in an
inverse power law temperature dependence of the total conductance {\it at
criticality}. 
%The transition, in the absence of tunneling, is robust and will 
%manifest itself in specific heat and charge susceptibility measurements.
%Tunneling,on the other hand, being a relevant operator sets a scale below
% which the system
%behaves in a non singular way, but for a wide range of parameters
%power law behavior of the  conductance will be observed. These
%observations are experimentally accessible.

In the related system of two {\it isotropically} coupled Kondo impurities, a 
quantum phase transition was predicted also, as a function of the interaction.
The results of numerical renormalization group studies\cite{jones} were 
confirmed by conformal field theoretical methods\cite{affleck}, and 
rationalized by phase shift arguments\cite{millis}. Unlike our case, the
same Fermi sea of electrons interacts with both impurites and Particle-Hole
symmetry is required to protect the fixed point. This may render it harder
to realize experimentally.

Let us start by considering a structure of two quantum dots, each coupled 
to their own leads. The lead-dot barriers are assumed to be narrow enough 
such that the tunneling is correctly modelled as a point contact. 
Furthermore we assume the presence of a strong enough magnetic field 
to achieve a fully spin-polarized electron gas. Thus the number of 
``flavors'', i.e. of the additional quantum numbers of transverse momenta
and spin of the electrons is restricted to one.
The dot is assumed to be relatively large, thus supporting a degenerate 
electron gas with small level spacing. This level spacing serves as a low 
energy cutoff, below which our scaling arguments do not hold. 
The Hamiltonian of one lead-dot system can then be written:
\begin{equation}
H= \sum_{k\alpha} \epsilon_{k} c_{k\alpha}^{\dagger}c_{k\alpha}+
J\sum_{kk'\alpha\ne\beta}c_{k\alpha}^{\dagger}c_{k'\beta} 
+ h.c.
\label{H}
\end{equation}
where $\epsilon_{k}$ is the energy of the electrons, $J$ is the tunneling
amplitude and the indices $\alpha$ and $\beta$ take the values $1$, when
referring to the lead and $2$ when describing electrons in the dot.
In a pseudospin notation the tunneling term is proportional
to $\sigma^+_{\alpha\beta} + \sigma^-_{\alpha\beta}$\cite{matveev}.

Next we recall that for small enough dots the Coulomb repulsion introduces
an interaction term between dot-electrons, the scale of which is 
$E_{C}= e^2/2C$, the charging energy, where $C$ is the capacitance of 
the dot. Experimentally it is also possible to tune
the overall potential of the system by a gate voltage $V_{\rm G}$. The
electrostatic energy of the dot can then be expressed as 
$E_{Q}=(Q-Q_{\rm G})^2/2C$, where (essentially)
$Q_{\rm G}=CV_{\rm G}$ and $Q$ is the charge on the dot. 
Tuning $Q_{\rm G}$ beyond $e/2$ makes it energetically favorable to transfer 
electrons across the barrier, giving rise to
the well-known set of parabolae as the ``band-structure'' of the system.
Transport becomes possible when the energies of states with different
number of electrons are degenerate. Thus the conductance shows sharp peaks
as a function of $Q_{\rm G}$, with maxima at $Q_{\rm G}/e= n+1/2$. In the vicinity of 
these degeneracy points the energies of the states with $n$ and $n+1$
electrons are much closer to each other than to any other state. It 
is then reasonable to truncate the Hilbert space to two states. 
A {\it second} pseudospin of $S=1/2$ can be introduced to 
represent this constraint on the allowed states. With this notation 
$H$ assumes the Kondo type form:
\begin{eqnarray}
H^{\rm K}= & &\sum_{k\alpha} \epsilon_{k} c_{k\alpha}^{\dagger}
c_{k\alpha}^{\phantom{\dagger}}\\
\nonumber
 + & & J \sum_{kk'\alpha\beta}c_{k\alpha}^{\dagger}
(\sigma_{\alpha\beta}^+ S^- + \sigma_{\alpha\beta}^- S^+ ) 
c_{k'\beta}^{\phantom{\dagger}}- \Delta S^z ~~.
\label{H2}
\end{eqnarray}
Here $\Delta$ is the gap between the $n$ and $n+1$ electron states
on the dot. The introduction of the two types of pseudospin
operators allows a complete mapping of a single dot to
the Kondo problem in a magnetic field $\Delta$, 
as first realized in its entirety by Matveev\cite{matveev}. 
Note that the Kondo term is {\it not} spin-rotationally invariant, 
it contains only the spin flip terms.

We proceed by including the interaction between the two dots
caused by their mutual capacitance $C_m$\cite{matveev2,halperin}.
The generated dot-dot coupling is proportional to $n_{\rm L}n_{\rm R}$, 
where we introduced the ${\rm L,R}$ notation
for the left and right dot respectively. Here $n_{\rm L,R}$ denote the charge
of the left or right dot. In pseudospin notation 
$S^z_{\rm L,R}=n_{\rm L,R}-1/2$.
The mutual capacitance in pseudospin notation gives rise to an
antiferromagnetic Ising type coupling:
$H_{\rm LR}^{\rm AF}= I_{z} S^z_{\rm L}S^z_{\rm R}$, 
where $I_{z} \sim E_{C_{m}}$.
The total Hamiltonian then takes the form:
$H = H_{\rm L}^{\rm K} + H_{\rm R}^{\rm K} + H_{\rm LR}^{\rm AF} $,
describing two anisotropic Kondo impurities, coupled by an antiferromagnetic 
Ising term. 

The physical content of the model can be analyzed by generalizing the 
arguments of the theory of the two-impurity Kondo model 
At $I_{z}=0$ we have two decoupled Kondo models. At T=0 in the magnetic 
language two independent {\it isotropic} Kondo singlets are formed with a
``binding energy'' $\sim T_{\rm K}$, as the anisotropy of the Kondo coupling is 
known to be irrelevant around this fixed point. 
In the charge language, the electrons form strongly hybridized 
states between the lead and the dot. This hybridization manifests itself by 
the strong coupling Kondo phase shift, $\delta=\pi/2$.
The key observation is that the ground state is a {\it singlet}.
At finite but small $I_z$ we follow Nozieres, who showed that
all operators around this fixed point break the singlet and thus are of 
dimension two. In particular the dot-dot interaction involves virtual hopping 
operators to fourth order in the lead-dot hybridization amplitude. 
The corresponding diagrams contain a large number of fermionic operators, 
and thus are irrelevant. Alternatively the large number of fermion operators
strongly confine the relevant phase space, leading to a positive
exponent for the temperature dependence. This consideration again yields a 
vanishing contribution at T=0.

In the opposite limit, $I_{z}=\infty$, the dot spins are aligned 
antiferromagnetically. The ($\uparrow ,\downarrow$) and 
($\downarrow ,\uparrow$) states are degenerate and form a {\it doublet}, 
which is independent of the conduction electrons. In the charge language 
these states consist of one extra electron being either on the left or on the 
right dot: $(1,0)$ and $(0,1)$. The energy of forcing on or taking away a 
dot-electron is $\sim I_{z}$, and is thus prohibited in this limit. 
Let us recall that the interaction term between the lead electrons and the 
pseudospin contains only spin raising and lowering terms. In the allowed 
Hilbert space the matrix elements of this coupling are zero. Therefore the 
phase shift of the conduction electrons vanishes. The degeneracy of the ground
state extends to large but finite $I_{z}$ couplings as well, since the flip 
from one state to another again requires a high power of fermion 
operators and thus is irrelevant. In other words, whichever dot the electron 
resides on, it hybridizes only with its corresponding lead. Since the dot-
dot coupling does not allow for charge transfer, these fluctuations remain 
confined to the dot and lead on the same side. Therefore the energies of the 
doublet's two states renormalize symmetrically for finite $I_z$, and thus the 
degeneracy is preserved. 

To sum it up, the ground state is a singlet for small values of $I_z$,
but changes its symmetry to a doublet at large values of $I_z$.
This change cannot be continuous: the two regions are  necessarily
separated by a phase transition. 
While this transition seems to be related
to the case of two Kondo impurities, it is certainly different and its
nature is as yet unexplored. In the remaining of this paper we assume
that the transition is of second order.

Here we pause to make connection to previous work by reviewing 
the band structure. The parabolae now have two indices, representing the 
charge states of the two dots. For $I_z=0$ the $(0,0)$ and $(1,1)$ curves
are touching $E_{Q}=0$, the latter displaced along the gate charge ($Q_{\rm G}$) 
axis by $e$. The $(0,1)$ and $(1,0)$ curves are centered at $Q_{\rm G}=e/2$, and 
are also shifted upward such that they go through the intersection of the 
$(0,0)$ and $(1,1)$ curves. Exactly this degeneracy of states with different
number of charges allows for transport across the dots, and thus gives
rise to the conductance peak. If one now introduces the 
mutual capacitance $I_{z}$, the upper parabolae are customarily shifted down 
by an amount $\sim I_{z}$. This creates two degeneracy points at 
$Q_{\rm G} \sim e/2 (1 \pm I_{z}/E_{C})$. Thus the original degeneracy of e.g. 
the $(0,0)$ and $(1,0)$ states, which allowed for the Kondo effect and
underlied much of the above considerations, seems to have been destroyed.
One might expect that whatever is left from the above picture, will
be observable at the shifted degeneracy points.

In contrast we predict that this new quantum critical point 
{\it is observable}, located precisely at the original value of $Q_{\rm G}=e/2$.
The reason for this is that for small $I_{z}$ the Kondo energy scale $T_{\rm K}$
is bigger than $I_{z}$. Therefore it is incorrect to construct a band
structure first and then try to include the Kondo physics. Instead
one has to {\it start} by accounting for the formation of the Kondo singlet,
a deeply non-perturbative effect. The subsequent inclusion of $I_{z}$ then 
means only a small perturbation, similar to a fluctuating magnetic field. 
According to the above reasoning such a field has a vanishing polarizing effect
on the Kondo singlet unless its (Zeeman) energy is comparable to $T_{\rm K}$. 
%In fact, according to the above 
%reasoning, the effect of a small $I_{z}$ is precisely zero.
Thus for $I_{z} < T_{\rm K}$ the $(0,1)$ and $(1,0)$ parabolae should not
be viewed as shifted from their $I_{z}=0$ location, and their degeneracy
is preserved. A strongly analogous situation occurs in single dots:
as shown first in Ref.\cite{guinea}, the effect of the Kondo type
many body processes is to strongly {\it collapse} the band structure,
sustaining their degeneracy up to some finite $I_{z}$.
On the other hand, for $I_{z} > T_{\rm K}$ it {\it is} a reasonable starting 
point to account for $I_{z}$ first, and then treat the Kondo coupling, i.e. 
the tunneling term as a perturbation. The two regimes are separated by the 
quantum critical point at$I_{z}^c \sim T_{\rm K}$.

Next we determine the total conductance of the two-dot structure. 
%So far tunneling between the two dots was expressly neglected.
The interdot tunneling Hamiltonian takes the form:
\begin{equation}
H_{tun}=I_{\pm} \sum_{kk'} 
c_{k{\rm L}2}^{\dagger} c_{k'{\rm R}2}^{\phantom{\dagger}} S_{\rm L}^{+}S_{\rm R}^{-}
+ h.c.
\label{tunnel}
\end{equation}
The pseudospin index $2$ appears explicitly, as we are considering dot-dot
tunneling. This term breaks time reversal symmetry,
and is a relevant perturbation at the quantum critical point.
One then expects that the low temperature behaviour of the renormalized
tunneling {\it at criticality} exhibits a singularity: 
$I_{\pm}(T) \sim T^{-\gamma}$. 
The conducting path across the structure involves a lead-dot, dot-dot
and dot-lead transitions. In the $I_{\pm} \ll J$ limit
the bottleneck, and thus the determining factor in the total conductance
$G$ is the dot-dot tunneling:
\begin{equation}
G(I_{z}=I_{z}^c,T) \sim (I_{\pm}(T))^2 \sim T^{-2\gamma} ~~ .
\label{G}
\end{equation}
This is only a crossover behaviour. As $T$ is further lowered, $I_{\pm}$ grows
large and flows to an attractive fixed point, controlling its asymptotic 
behavior.
Given its analogous structure, it is plausible that the dimension of 
$I_{\pm}$ is the same as that of the particle-hole symmetry breaking operator.
However the actual value of $\gamma$ still needs to be determined ~\cite{ani}.
%In the one-channel two impurity problem with spin rotational symmetry 
%this dimension is $\gamma=1/2$ \cite{jones}, but it is not clear if
%spin anisotropic model scales to the same fixed point.
%In the two-channel case, realized if the magnetic field is 
%switched off or two transverse momenta are allowed by the experimental
%layout, it can be shown \cite{ani} that the Ising term 
% is marginal around the ``decoupled'' fixed point, leading
% to a line of  fixed points which terminates at some intermediate value.

What happens away from criticality?
For $I_{z}<I_{z}^c$ the Kondo singlets inhibit the transport.
At $T=0$ the binding is complete, thus $G(T=0)=0$.
Concentrating once again on the bottleneck dot-dot tunneling we compute 
the scaling dimensions of the involved operators. The fermion operators
carry dimension 1/2, the spin raising operator has dimension 1.
The current operator is constructed from the $[N,H]$ commutator.
Collecting the terms the current-current correlator decays with the 
sixth power of time. Substituting this into the Kubo formula finally 
yields $G(T) \sim T^4$. The lead-dot process occurs via the Kondo coupling
which scaled to its unitarity limit, thus it does not give rise to additional
powers ot $T$. 
%To determine the conductance at finite temperatures
%one recalls that the Kondo singlet is not bound by an energy gap, it is only
%a strongly correlated state with a characteristic energy $T_{\rm K}$.
%External perturbations are thus capable of polarizing it effectively. 
%This feature is reflected in the Fermi-liquid like behaviour
%of the response functions, exhibiting simple integer power-law 
%$T$ dependences. In particular the electron lifetime behaves as 
%$1/\tau \sim T^2$. This sets the attempt frequency for tunneling, 
%thus the {\it conductance} $G(T)$ will be proportional to $T^2$  \cite{note}.

In the regime $I_{z}>I_{z}^c$ electrons have to break an
Ising bond.
Thus at zero temperature again $G(0)=0$, and at finite $T$
the temperature dependence takes an activated form,
$G(T) \sim \exp(-W/T)$, where $W \sim I_{z}$. To sum it up, the 
conductance as a function of $I_{z}$ at zero temperature is zero nearly 
everywhere, and exhibits a resolution limited peak {\it at} $I_{z}=I_{z}^c$.
At finite but low temperatures the peak persists in $G(T)$ as a function
of $I_{z}$. On the two sides of the peak $G$ assumes non-zero values.
These wings are asymmetric, with $T$ dependent values.
The different regimes are shown in Fig.1.

\begin{figure}
\epsfxsize=3.0in
\epsfysize=2.25in
\epsffile{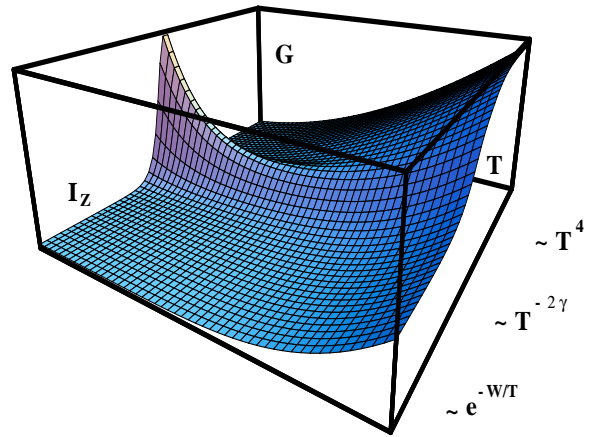}
\vskip 0.2cm
\caption{The conductance $G$ as a function of T and $I_{z}$.}
\end{figure}

Finally we examine the effect of changing the gate voltage, which tunes
the gate charge away from the special point $Q_{\rm G}=e/2$, considered so far.
In the Kondo language this gives a finite value to the magnetic field $\Delta$.
For $I_{z}<I_{z}^c$ we utilize the same observation as before: for 
the unperturbed state is the Kondo singlet. Thus 
for $\Delta = 0 $, i.e. {\it at} $Q_{\rm G}=e/2 $ transport is still impossible
at $T=0$. For $0<\Delta < T_{\rm K}$ the singlet is somewhat polarized, and
weak transport is possible. This manifests itself in two small-amplitude 
``shadow-bands'' in a V shape determined by $|\Delta|=I_{z}$.
This is the location of the crossing of the typically constructed ``shifted
parabolae''. An important transport channel in this region is co-tunneling,
which only virtually breaks the Kondo singlet.
For $I_{z}=I_{z}^c$ the pronounced conductance peak of the quantum critical 
point is present {\it at} $\Delta = 0$. This peak continues out to finite 
$\Delta$, forming a parabola-like ridge, which smoothly connects to the usual 
split conductance peaks at $|\Delta|=I_{z}$ for $I_{z}>I_{z}^c$. In this 
region $I_{z}$ is the larger of the energy scales and constructing the band 
structure first is appropriate.

Constructing the picture from the large $I_{z}$ side, the 
magnetic field $\Delta$ is trying to induce a spin-flip transition in the 
antiferromagnetic singlet. It is competing with the singlet binding energy, 
so the spin-flip can only occur when the binding energy equals the Zeeman 
energy: $|\Delta| \sim I_{z}$, forming the usual V locus for the split peaks. 
Approaching the quantum critical point however the binding energy 
{\it collapses to zero}, hence the V 
becomes rounded, and closes at $I_{z}^c$, as shown in Fig.2.

\begin{figure}
\epsfxsize=3.0in
\epsfysize=2.25in
\epsffile{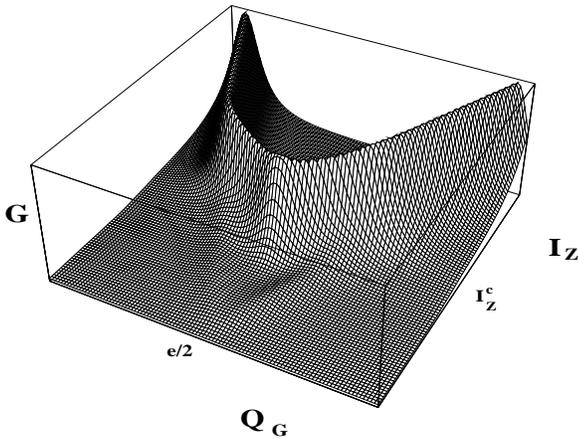}
\vskip 0.1cm
\caption{The conductance peak as a function of $I_{z}$ and the gate voltage
at $T=0$.}
\end{figure}

To summarize, the key experimental predictions of our calculations are:\\
{\it (i)} If the gate voltage is fixed so that $Q_{\rm G}=e/2$, then by tuning
$I_{z}$ a pronounced new conductance peak has to be observed at some critical
value $I_{z} = I_{z}^c \sim T_{\rm K}$.\\
{\it (ii)} Staying at this point $Q_{\rm G}=e/2, I_{z} = I_{z}^c$, the conductance
$G(T)$ should exhibit a power-law singularity in its temperature dependence.\\
{\it (iii)} The amplitude of the split conductance peaks at $Q \ne e/2$
should exhibit a marked collapse as a function of $I_{z}$ when
$I_{z}$ approaches $I_{z}^c$ from above.

We know of no experimental observation of these predictions yet.
This maybe due to the fact that the above theory applies only under the 
following conditions:\\
%{\it i)} The tunneling barrier also acts as a potential scatterer. This
%breaks particle-hole symmetry and may obliterate the quantum fixed point.
%One can avoid this by chosing a smooth profile for the constriction
%and adjust it to have the possible highest transmission coefficient.\\
{\it (i)} Typical experiments \cite{westerwelt} and the corresponding 
theory \cite{matveev2} considered the case of fixed $I_{z}$, and described
the evolution of the peak structure with tuning of $I_{\pm}$, the dot-dot
tunneling. The present theory addresses the case of fixed and small
$I_{\pm}$, and tuning with $I_{z}$ instead.\\
{\it (ii)} The number of tunneling channels should be small.

The above theory strictly applies only for the case of a single channel.
This requires a narrow, long constriction between the leads and the dot, 
similar to the case considered in \cite{matveev3}. We expect important changes
when the number of flavors of the electrons is increased.
Switching off the magnetic field increases the number of channels to two.
It can be shown \cite{ani} that the Ising term 
is marginal around the ``decoupled'' fixed point, leading
to a line of  fixed points which terminates at some intermediate value.
The fixed point structure of the related two Kondo impurities
model changes analogously.
The Kondo singlet and the antiferromagnetic singlet phases remain
intact, but the quantum critical fixed point expands into a
very unusual {\it fixed point area}, a whole region of the parameter space
consisting of fixed points\cite{jones2}. If such a structure emerges in
our case, then a broadened conductance
peak will form as a function of $I_{z}$ at $T=0$, and the finite temperature
conductance should exhibit singular temperature dependence with $I_{z}$ 
dependent exponents. 
The case of even larger number of channels has been
investigated for single scatterers in relation to the physics of
two level systems\cite{vladar}. It has been shown that a two dimensional 
subspace of the flavor indices emerges to dominate exponentially over the 
others in the course of scaling. Therefore we expect the basic
picture of two distinct phases and a well defined quantum phase transition 
in between to carry over, but obviously further calculations are needed
on this point. 

In sum we studied the system of two coupled quantum dots. We established
the existence of an intriguing new quantum critical point. 
Several experimentally accessible predictions were reached: a new conductance 
peak at $Q_{\rm G}=e/2$, an inverse power law dependence of the conductivity at 
this same point and a marked collapse of the split conductance peaks,
when the experimental parameters are in the suitable range.

We enjoyed illuminating discussions with I. Affleck, A. Georges
B. Jones, A. Ludwig, A. Millis,  T. Pohjola, A. Sengupta and G. Zar\'and.
This work has been supported by NSF-DMR-95-28535 and the SFB 195 of
the DFG.

\end{document}